\documentstyle[preprint,prb,aps]{revtex}
\begin{document}
\draft
\title{A Scanned Perturbation Technique For Imaging Electromagnetic Standing Wave
Patterns of Microwave Cavities}
\author{Ali Gokirmak, Dong-Ho Wu, J. S. A. Bridgewater, and Steven M. Anlage}
\address{Center for Superconductivity Research, Department of Physics, University of\\
Maryland, College Park, MD 20742-4111 }
\maketitle

\begin{abstract}
We have developed a method to measure the electric field standing wave
distributions in a microwave resonator using a scanned perturbation
technique. Fast and reliable solutions to the Helmholtz equation (and to the
Schr\"{o}dinger equation for two dimensional systems) with
arbitrarily-shaped boundaries are obtained. We use a pin perturbation to
image primarily the microwave electric field amplitude, and we demonstrate
the ability to image broken time-reversal symmetry standing wave patterns
produced with a magnetized ferrite in the cavity. The whole cavity,
including areas very close to the walls, can be imaged using this technique
with high spatial resolution over a broad range of frequencies.
\end{abstract}


\section{Introduction}

Quantitative mapping of the electromagnetic fields inside a metallic cavity
is important for various technological and scientific reasons. An
experimental understanding of the behavior of electromagnetic waves in
irregularly shaped structures is important for the design of electrical
circuits, particularly those housed inside metallic containers which are
susceptible to strong concentrations of electromagnetic fields. The degree
of concentration of the field depends sensitively on the operating frequency
and shape of the device. Of more scientific interest, electromagnetic
resonators can be used to represent quantum mechanical potential wells. In
this case, the standing wave patterns can reveal the probability density of
the corresponding solution to the Schr\"{o}dinger equation for an infinite
square well potential of arbitrary shape. These eigenfunctions are important
for understanding the behavior of mesoscopic structures, and will be crucial
for the design of nano-scale electronic devices. In addition, a great deal
of interesting physics can be explored by means of an experimental
understanding of the behavior of the wavefunctions in irregular shaped
devices, including wave localization and wavefunction fluctuations. To
investigate the wavefunctions, one needs a simple and reliable method of
imaging electromagnetic waves in cavities. In this paper we present a simple
method to image the standing wave eigenfunctions of two-dimensional
electromagnetic cavities, and their corresponding quantum eigenfunctions.

Imaging of electromagnetic standing waves in microwave resonators was
pioneered by Slater and coworkers.\cite{Slater} They used the technique of
cavity perturbation to selectively measure the microwave electric and
magnetic field amplitudes using a scanned perturbation. Others have refined
this technique by modulating the perturbation to measure both electric and
magnetic fields.\cite{Rebeiz} More recently, forms of near-field ``microwave
microscopy'' have been developed to probe electromagnetic fields on small
length scales (much less than the free space wavelength of the radiation) in
resonant \cite{Ashfaq} and non-resonant \cite{Bloom,vdW,Anlage} devices.
Other recent derivatives of Slater's work have examined the standing wave
patterns of electromagnetic resonators which simulate wave chaotic systems
in two dimensions. \cite{Sridhar1,Sridhar2,Stein+Stockmann,Wu} In several of
the techniques mentioned above, one measures a mixture of electric and
magnetic fields, and some methods are not easily scaled to high frequency
imaging. Further none of these methods has demonstrated the ability to image
eigenfunctions in the absence of time reversal symmetry, which is an
important consequence of device performance under magnetic field. We have
developed an experimental method of imaging primarily the microwave electric
field amplitude which operates over a very broad range of frequencies with
or without time-reversal symmetry. In this paper we describe the
experimental method along with our experimental results obtained from quasi
two-dimensional cavities, both integrable and nonintegrable. The method can
be used to test fundamental wave dynamics in an irregularly shaped
structure, and also can be used for performance tests of cavity-based
microwave devices.

For the investigation of fundamental wave (or quantum) mechanical behavior,
one exploits the electromagnetic analog \cite{Hupert1,Hupert2} of the
Schr\"{o}dinger equation in two dimensions (2D). Both the time-independent
Schr\"{o}dinger equation and the Helmholtz equation are linear second order
differential equations with respect to space with constant coefficients:

\begin{equation}
\nabla ^2\Psi _n+2m\left( \epsilon _n-V\right) /\hbar ^2\Psi _n=0
\label{Schrodinger Eq.}
\end{equation}

\begin{equation}
\nabla ^2E_{zi}+k_i^2E_{zi}=0  \label{Helmholtz Eq.}
\end{equation}

The Schr\"{o}dinger equation, Eq. \ref{Schrodinger Eq.}, describes the n$%
^{th}$ excited eigenfunction $\Psi _n$ of a particle of mass m in a
potential V, where $\epsilon _n$ is the energy of the n$^{th}$ excited
state, and $\hbar $ is Planck's constant divided by 2$\pi $. In the
Helmholtz equation, Eq. \ref{Helmholtz Eq.}, $E_{zi}$ is the z-component of
the electric field, and $k_i$ is the wave vector in the propagation
direction of the i$^{th}$ mode (it is assumed that the resonator is much
smaller in the z-direction than in the x and y directions). In general $%
k_i^2 $ = $\omega _i^2\mu \epsilon $, where $\omega _i$ is the i$^{th}$
resonant frequency of the cavity with uniform permittivity $\varepsilon $
and permeability $\mu $. The probability density 
%TCIMACRO{\TEXTsymbol{\vert}}
%BeginExpansion
\mbox{$\vert$}%
%EndExpansion
$\Psi _n$%
%TCIMACRO{\TEXTsymbol{\vert}}
%BeginExpansion
\mbox{$\vert$}%
%EndExpansion
$^2$ in the Schr\"{o}dinger equation for the quantum mechanical problem is
analogous to 
%TCIMACRO{\TEXTsymbol{\vert}}
%BeginExpansion
\mbox{$\vert$}%
%EndExpansion
$E_{zi}$%
%TCIMACRO{\TEXTsymbol{\vert}}
%BeginExpansion
\mbox{$\vert$}%
%EndExpansion
$^2$ in the Helmholtz equation for the electromagnetic problem, as long as
the $2m\left( \epsilon _n-V\right) /\hbar ^2$ and $k_i^2$ terms in these
equations are constant. In a thin microwave cavity where $E_z$ is uniform in
the z-direction, the solutions to the Helmholtz equation with perfectly
conducting walls ($E_{zi}$ = 0 at the boundary) are equivalent to the
solutions of the two-dimensional Schr\"{o}dinger equation with hard wall
boundaries ($\Psi _n$ = 0 at the boundary) of the same geometry.

The experimental technique we developed uses a perturbation to obtain a
spatially resolved measure of 
%TCIMACRO{\TEXTsymbol{\vert}}
%BeginExpansion
\mbox{$\vert$}%
%EndExpansion
$E_{zi}$%
%TCIMACRO{\TEXTsymbol{\vert}}
%BeginExpansion
\mbox{$\vert$}%
%EndExpansion
$^2$ in a hollow microwave cavity. At the same time, the technique allows us
to measure the analogous quantum eigenfunctions of 2D infinite square well
potentials. We can use this method to study the properties of any 2D quantum
dot where the solution to Schr\"{o}dinger equation is needed, and analog
solutions for the eigenvalues and eigenfunctions can be obtained.

The outline for the rest of the paper is as follows. In section II we
discuss the principle of the measurement in detail, while in section III we
describe the experimental setup. In section IV we show some sample data, and
discuss its general features and make comparisons with theory. The
limitations and constraints of the technique are discussed in section V,
while in Section VI we summarize our findings.

\section{Principle of the measurement}

When microwave electromagnetic fields in a cavity are perturbed, the
resonant frequency $\omega $ is altered as,\cite{Slater}

\begin{equation}
\omega ^2=\omega _0^2\left( 1+\int \left( B^2-E^2\right) dV_p\right)
\label{Slater Eq.}
\end{equation}

where $\omega _0$ is the resonant frequency of the unperturbed cavity, B is
the microwave magnetic field and E is the microwave electric field at the
location of the perturbation. The integral is taken over the perturbation
volume $V_p$. The field components in the expression are normalized over the
volume of the cavity such that: $\int E^2dV=1$, $\int B^2dV=1$. If we change
the location of a metallic perturbation in the cavity and record the
resonant frequency for each point, we can image a combination of $E^2$ and $%
B^2$ inside the cavity. Note that while the contribution of $E^2$ results in
a frequency shift to a lower frequency (due to the negative sign in front of 
$E^2$), the contribution of $B^2$ results in a shift to a higher frequency.
Throughout the development of our experimental technique, we find that the
shape of the perturbation determines the relative contribution of $E^2$ and $%
B^2$ to the shift in the resonant frequency.

Both pins and spheres are used as perturbations in the measurements
presented here. Spheres introduce significant contributions to the frequency
shift from the magnetic field components. In our measurements we want to
image only 
%TCIMACRO{\TEXTsymbol{\vert}}
%BeginExpansion
\mbox{$\vert$}%
%EndExpansion
$E_z$%
%TCIMACRO{\TEXTsymbol{\vert}}
%BeginExpansion
\mbox{$\vert$}%
%EndExpansion
$^2$ in the cavity, hence pin perturbations are used to minimize the
magnetic field contribution .

The electromagnetic modes of a two-dimensional microwave cavity are
particularly simple, as there are only 3 non-zero components of the
electromagnetic field. A 2D cavity is designed to have a height (z)
significantly smaller than the dimensions in the horizontal plane (x, y) so
that there is no wavevector component in the z-direction below a
well-defined cutoff frequency. For our cavities with a height of 0.310'',
below 19.05 GHz only 2D transverse magnetic (TM) modes can propagate, hence
only $E_z$, $B_x$, and $B_y$ can have nonzero values (see Fig. \ref{Bow-Tie
Cavity}). There are approximately 10$^3$ TM modes below 19.05 GHz in the
quarter bow-tie cavity discussed below (Fig. \ref{Bow-Tie Cavity}).\cite
{PaulSo}

\section{Experimental setup}

As shown in Fig. \ref{Schematic}, the cavity is imaged using a microwave
vector network analyzer, and a two-dimensional scanner which moves a small
metallic perturbation inside the cavity through the influence of an external
magnetic field. The perturbation can be inserted and removed through a small
coupling hole on the cavity lid so that the cavity does not have to be
opened each time the perturbation is changed. In the rest of this section we
shall discuss the types of microwave cavities we have imaged, the microwave
system for the measurement, the perturbation scanner and the perturbations
themselves.

\subsection{Cavity}

In our experiment, a classically chaotic two-dimensional potential well is
used. Four intersecting circles with radii smaller than the separation
between the centers form a closed region resembling a bow tie. The system we
use in our experiments is one quarter of this region (see Fig. \ref{Bow-Tie
Cavity}). This geometry ensures that all typical ray-trajectory orbits are
chaotic and all periodic orbits are isolated,\cite{PaulSo,Antonsen} and
allows one to study the eigenmodes of wave chaotic systems.\cite{Heller}

The quarter bow tie is carved into brass and the surface is plated with
copper to reduce the surface resistance and increase the quality factor of
the cavity modes. A copper-plated flat lid closes the cavity from the top.
There are four holes on the lid where the coupling probes are inserted into
the cavity. The quality factor (Q) for the bow-tie cavity range from 700 to
3000, depending upon the resonance.

A second rectangular cavity was used for measurements in a simple geometry
to check the validity of our experimental technique. The dimensions of the
rectangular cavity are 7.5'' $\times $ 14.0'' $\times $ 0.310'' high.

\subsection{Vector Network Analyzer and the Microwave Setup}

An HP8722D vector network analyzer (NWA) was used to measure the resonant
frequency of the cavity for each position of the perturbation. A signal
coming out of the network analyzer is amplified by an HP8349B microwave
amplifier (see Fig. \ref{Schematic}). The output of the amplifier is
injected into the cavity through a E-plane coupling probe, and the
transmitted signal from the cavity is picked up by a second E-plane coupling
probe (see Figs. \ref{Schematic} and \ref{Magnet Ensemble}). The coupling
probes are made out of semi-rigid coaxial cable by stripping the outer
conductor off of one end, with the length of the exposed inner conductor
slightly less than the cavity height. The coupling probes are connected to
the top lid of the cavity with adjustable micrometers and electrically
isolated from the cavity. The signal picked up by the second coupling probe
is taken to the detector of the network analyzer. The resonant frequencies
of the cavity correspond to local maxima in the transmission amplitude
between the two ports of the network analyzer in the frequency domain.

\subsection{Scanning Setup}

A two-dimensional scanner\cite{Arrick} carrying a magnet ensemble under the
cavity is used to move a perturbation inside the cavity to map out the
electric field. The ensemble is composed of two magnets and a magnet iron
cone on top (see Fig. \ref{Magnet Ensemble}). To produce an isotropic and
strong restoring force on the steel perturbation inside the cavity, the
field from the magnets is focused by the magnet iron cone. The platform is
moved on steel rails by means of belts driven by stepper motors. In order to
image the fine details of high frequency eigenmodes, we redesigned the
scanner to have as little static and dynamic friction as possible between
the carriage and the rails supporting it. This significantly increased the
accuracy in the magnet location compared to the original design of the
scanner.

Further, to reduce friction and keep the magnet ensemble at a constant
separation from the cavity, a thin Teflon piece was inserted between the
cone tip and the cavity. The magnet and cone combination is supported by
four springs glued onto the mobile platform. The magnets and cone are placed
inside a square case and are free to move up and down to follow the small
changes in the scanner-cavity separation. The case is made out of steel to
concentrate the return magnetic flux from the cone. This minimizes the
effect of the scanning magnets on magnetically sensitive objects in or
around the cavity, which are unrelated to the perturbation (e.g. a ferrite
bar introduced into the cavity for some of the measurements).

\subsection{Data acquisition}

A computer controls both the network analyzer and the scanner and records
the data transferred from the network analyzer. The network analyzer
measures the transmission magnitude through the cavity and is set to take
data at 1601 frequency points per span, with a typical span of 3 MHz
centered on one of the resonant modes. Data points on the network analyzer
are smoothed to reduce the noise level in the data. The resonant peak is
followed as the perturbation is moved, and the frequency corresponding to
the maximum transmission amplitude is recorded at each stationary position
of the perturbation. The perturbation is moved on a square grid with step
sizes ranging from 0.2'' to 0.05'', depending on the frequency of the mode
to be imaged (there are 71600 data points on the images made with a step
size of 0.05''). For most of the perturbations, we observed that the maximum
shift in resonant frequency due to the perturbation is on the order of 10
MHz. To test the resolution for frequency shift, we examined a rectangular
cavity eigenmode of resonant frequency 1.5 GHz. The 3dB bandwidth of this
mode was 6.5 MHz (the quality factor of this mode is 1250) and we could
reliably distinguish a shift on the order of 5 kHz. Thus we are able to
distinguish changes in the resonant frequency of about one part in 10$^3$ of
the 3 dB bandwidth.

\subsection{Perturbation}

In the development of this technique, we have tried various sizes of
metallic pins and spheres to find the optimum perturbation. Experiments
indicate that the perturbation can be significantly smaller than the
wavelength of the radiation and still produce excellent images. A
perturbation introduced into the cavity shifts the resonant frequency of the
cavity by an amount proportional to a combination of the electric and
magnetic field amplitudes at the location of perturbation. For a spherical
metallic perturbation, the magnetic field contribution to the shift in the
resonant frequency is half of the contribution of the electric field :\cite
{Slater}

\begin{equation}
\frac{\omega ^2-\omega _0^2}{\omega _0^2}=3\left( \frac{4\pi }3r_0^3\right)
\left( \frac 12B_0^2-E_0^2\right)  \label{Sphere Eq.}
\end{equation}

where $r_0$ is the radius of the spherical perturbation, $E_0$ is the
averaged electric field, $B_0$ is the averaged magnetic field, each being
separately normalized, and averaged over the volume of the perturbation. The
spectrum of frequency shifts for a spherical perturbation (radius of 1/16'')
in a rectangular cavity eigenmode shows the expected distribution (Fig. \ref
{Distributions}a). Approximately 30\% of the frequency shift data turns out
to be higher than the resonant frequency of the empty cavity. These are the
points where the magnetic field contribution dominates the electric field
contribution. However, the negative frequency shift is, at least partially,
reduced by the magnetic field contribution. For our purposes this is
unacceptable because we want to image only 
%TCIMACRO{\TEXTsymbol{\vert}}
%BeginExpansion
\mbox{$\vert$}%
%EndExpansion
$E_z$%
%TCIMACRO{\TEXTsymbol{\vert}}
%BeginExpansion
\mbox{$\vert$}%
%EndExpansion
$^2$.

To meet the goal of imaging electric field in the cavity, we use a pin with
rounded ends to scan inside the cavity. The spectrum of frequency shifts
produced by a pin perturbation is shown in Fig. \ref{Distributions}b. Since
most of the resonant frequencies are below the unperturbed resonant
frequency, it is clear that the pin measures mainly the electric fields in
the cavity. For frequencies below 9 GHz, we use a pin whose body has a
diameter 0.0220'' and height 0.2430''. For frequencies higher than 9 GHz we
use a smaller pin with cylindrical body diameter 0.0085'' and height
0.1535''. Experimentally, we observe about a 10\% contribution from the
magnetic field components for these pins (Fig. \ref{Distributions}b).

\subsection{Imaging With a Magnetized Ferrite in the Cavity}

It is of interest to investigate the eigenvalues and eigenfunctions of
chaotic quantum systems with and without time reversal symmetry.\cite
{Wu,PaulSo} By introducing a ferrite into the cavity and magnetizing it, one
can add a non-reciprocal phase shift to the electromagnetic waves in the
cavity, and thus break time-reversal symmetry.\cite{Wu,PaulSo,PaulSoPhD} The
off-diagonal terms in the ferrite permeability tensor induce a phase shift
for reflected waves off of the ferrite which changes magnitude significantly
when the time evolution of the wave is reversed. However, due to the complex
nature of ferrite electrodynamics, the amount of this non-reciprocal phase
shift is strongly frequency dependent,\cite{PaulSoPhD} and only becomes
large enough to fully break time reversal symmetry in a frequency window
above 13.5 GHz for the cavity shown in Fig. \ref{Bow-Tie Cavity}. Hence it
is imperative to image eigenfunctions up to the highest frequencies possible
to see the effects of time reversal symmetry breaking on the wave chaotic
eigenfunctions.

The ferrite used in the measurements\cite{TransTech} is 0.2'' thick, 0.310''
high, and 8.4'' long. It is placed adjacent to the short linear boundary on
the left side of Fig. \ref{Bow-Tie Cavity}. The ferrite is magnetized with
ten 2'' x 2'' x 0.5'' magnets placed outside the cavity. The magnets are
placed over and under the cavity, centered on the ferrite in two linear
arrays. These magnets provide a uniform DC magnetic field in the z
direction. The magnets are held in place with a C-shaped steel piece, which
concentrates the return magnetic field flux and minimizes the effect of
these stationary magnets on the perturbation scanned inside the cavity.
Similarly the scanner magnet is placed in a steel holder concentrating the
field and minimizing the effect of the scanning magnet on the ferrite.
Although the effect of the scanner magnet is reduced significantly with the
steel holder, there still is a visible change in the resonant frequency of
the cavity as a function of the scanner magnet location not caused by the
perturbation. This undesired effect is significantly reduced by subtracting
a background from the eigenmode image (see Section IV, part C).

\section{Data}

\subsection{Comparison Between Theory \& Experiment For Rectangular Wave
Functions}

We first examine the spectrum of frequency shifts produced by the sphere and
pin perturbations for a given mode of the rectangular resonator. It is clear
from Fig. \ref{Distributions} that a pin perturbation produces a more
faithful image of the electric fields inside the microwave cavity. Let us
now examine the degree to which the behavior of the pin and sphere
perturbations agree with Slater's analysis. We consider the rectangular
cavity mode images presented in Figs. \ref{Pin Image} (pin perturbation) and 
\ref{Ball Image} (sphere perturbation). The spectrum of frequency shifts for
these images are shown in Fig. \ref{Distributions}. The order of magnitude
maximum frequency shift caused by a spherical perturbation in a previously
uniform electric field is given by Eq. \ref{Sphere Eq.} and the radius of
the sphere given above. For the TM$_{170}$ mode shown in Fig. \ref{Ball
Image}, the theoretical result is $\left[ \left( \omega _{\min }^2-\omega
_0^2\right) /\omega _0^2\right] $ = -2.86 x 10$^{-4}$, which should be
compared to the experimental value $\left[ \left( \omega _{\min }^2-\omega
_0^2\right) /\omega _0^2\right] $ = -7.78 x 10$^{-4}$. The magnetic field
perturbation is predicted to produce a frequency shift of $\left[ \left(
\omega _{\max }^2-\omega _0^2\right) /\omega _0^2\right] $ = 1.32 x 10$^{-4}$
versus an observed value of $\left[ \left( \omega _{\max }^2-\omega
_0^2\right) /\omega _0^2\right] $ = 2.05 x 10$^{-4}$. In both cases, the
predicted frequency shifts are approximately a factor of 2 lower than the
observed values. Seen another way, the ratio of maximum electric to maximum
magnetic perturbation is predicted to be approximately 2.2, whereas we
observe a ratio of about 3.8. One reason for these discrepancies is that
this calculation assumes the sphere is placed into an initially uniform
electric field (i.e. in the middle of a parallel-plate capacitor with a
plate separation much greater than the perturbation diameter) However, in
our measurement the sphere lies in contact with one plate of the parallel
plate capacitor, thus significantly altering the field and producing a
larger perturbation.

For the pin perturbation, the data is in less agreement with theory,
although in a favorable direction for electric field imaging. For the TM$%
_{170}$ mode shown in Fig. \ref{Pin Image}, the electric field perturbation
is predicted\cite{Slater} to be $\left[ \left( \omega _{\min }^2-\omega
_0^2\right) /\omega _0^2\right] $ = -3.30 x 10$^{-4}$, compared to the
experimental value $\left[ \left( \omega _{\min }^2-\omega _0^2\right)
/\omega _0^2\right] $ = -1.8 x 10$^{-3}$. The magnetic field perturbation is
predicted to produce a frequency shift of $\left[ \left( \omega _{\max
}^2-\omega _0^2\right) /\omega _0^2\right] $ = 1.04 x 10$^{-5}$ versus an
observed value of $\left[ \left( \omega _{\max }^2-\omega _0^2\right)
/\omega _0^2\right] $ = 4.9 x 10$^{-5}$. The ratio of maximum electric to
maximum magnetic perturbation is predicted to be approximately 32, whereas
we observe a ratio of about 37, again indicating a stronger perturbation
which favors the electric fields. The reason for these discrepancies may
again be due to the fact that the pin is in electrical contact with one of
the walls of the resonator. We speculate that there is a ``lightning rod''
effect between the top of the pin and the top lid of the cavity which gives
rise to enhanced electric field perturbation with no additional contribution
from the magnetic field perturbation.

The rectangular cavity images in Figs. \ref{Pin Image} and \ref{Ball Image}
show the resonant frequency of the cavity as a function of the position of
the perturbation. The pin image (Fig. \ref{Pin Image}) shows a distribution
which closely resembles the electric field amplitude squared distribution in
a TM$_{170}$ standing wave of a rectangular resonator. The sphere image, on
the other hand shows an additional feature which is due to contribution of
the magnetic fields.

Line cuts of the data are shown on the sides of Figs. \ref{Pin Image} and 
\ref{Ball Image}. The upper line cut (b) is a vertical cut through the 7
main features, while the lower line cuts (c) are horizontal cuts through the
data. Also shown in these line cuts are the expected theoretical frequency
shifts based on solutions to the Helmholtz equation, calculated from
Slater's formulas and scaled to fit the dynamic range of the data. The
vertical line cut in the pin image (Fig. \ref{Pin Image}b) shows excellent
agreement with the expected sinusoidal modulation of the resonant frequency.
However, because of the additional magnetic field contributions, the
vertical line cut of the sphere image (Fig. \ref{Ball Image}b) can also be
fit to a simple sinusoidal modulation which incorrectly overestimates the
probability amplitude (dashed line fit in Fig. \ref{Ball Image}b). In this
case, to extract just the electric field variation, one must identify the
unperturbed resonant frequency in Fig. \ref{Distributions}a, and fit to a
sinusoidal squared deviation from that frequency (solid line fit in Fig. \ref
{Ball Image}b). Likewise, the horizontal line cut of the pin image (Fig. \ref
{Pin Image}c, line cut A) shows a simple sinusoidal modulation through the
peak, and no measurable modulation between the peaks (line cut B). The
corresponding horizontal line cuts of the sphere image (Fig. \ref{Ball Image}%
c) show sinusoidal modulation from electric field contribution (downward
deviation) and magnetic field contributions (upward deviation). This
analysis shows that the pin perturbation imaging method effectively
eliminates contributions from magnetic fields in the standing wave images of
our microwave resonator.

\subsection{Wave Chaotic Cavity Images}

Shown in Figs. \ref{Low Freq. Chaos Mode}, \ref{Mid Freq. Chaos Mode}, and 
\ref{High Freq. Chaos Mode} are probability amplitude 
%TCIMACRO{\TEXTsymbol{\vert}}
%BeginExpansion
\mbox{$\vert$}%
%EndExpansion
$\Psi _n$%
%TCIMACRO{\TEXTsymbol{\vert}}
%BeginExpansion
\mbox{$\vert$}%
%EndExpansion
$^2A$ images made with a pin perturbation of the bow-tie resonator at 3.46,
5.37, and 11.94 GHz.\cite{ColorPics} Here A is the area of the cavity, $\int
|\Psi _n|^2dA=1$, and the frequency shifts above the unperturbed value are
re-defined to be zero in the plots. The field distributions in the chaotic
eigenmode images are rather complicated and intricate. Although the patterns
seem to be irregular, the probability amplitude maxima often form local
regions with circular and linear structures. The linear structures, like
those in Fig. \ref{High Freq. Chaos Mode}, were noted by Heller in
wavefunctions produced by a random superposition of planes waves of fixed
momentum.\cite{Heller} Heller also noted the semi-circular congregations of
local maxima which surround regions of low probability amplitude, like those
evident in Figs. \ref{Mid Freq. Chaos Mode} and \ref{High Freq. Chaos Mode},
in random superpositions of plane waves. His explanation was that regions of
low probability amplitude must be the source of radial nodes in the
wavefunction, giving rise to semi-circular clusters of high probability
amplitude in the immediate surroundings.\cite{Heller}

The low frequency modes shown in Figs. \ref{Pin Image}, \ref{Ball Image},
and \ref{Low Freq. Chaos Mode} gives us an opportunity to investigate the
noise present in our images. From analysis of rectangular eigenfunctions
where the wavefunctions are known, we can measure the signal-to-noise ratio
(SNR) of the images. The SNR is defined as the ratio of the frequency shift
variation to the root mean square deviation between the predicted and
observed frequency shift for a rectangular wavefunction. We find that SNR is
in the range of 30 to 60 at high input powers (+20 dBm from the amplifier)
and does not degrade significantly even at the lowest source powers we used
(-18 dBm). We thus have a very robust and clear method of imaging
eigenfunctions.

The mode shown in Fig. \ref{High Freq. Chaos Mode} has features which are on
the order of 0.5'' in size. We find that features as small as 0.25'' can be
resolved by our imaging technique. The highest frequency mode imaged to date
is at 15.4 GHz, where the guided wavelength in the resonator is
approximately 1 inch. In principle one could image at even higher
frequencies, however the density of modes becomes too great to image without
mode mixing.

A detailed analysis of the chaotic eigenfunctions is done through
statistical analysis of their properties. The probability amplitude
distribution function and the two-point correlation function of the
eigenfunctions are important properties which are sensitive to the
integrability of the potential well, and the presence or absence of
time-reversal symmetry. These properties are discussed in detail elsewhere.%
\cite{Wu,Prigodin}

\subsection{Background Subtraction}

When imaging broken time-reversal symmetry eigenstates with a ferrite
present in the cavity, the scanning magnets have a slight effect on the
ferrite due to the variable magnetic flux seen by the ferrite. This causes a
background shift in the resonant frequency of the cavity mainly dependent on
the scanning magnet location on the x-axis (Fig. \ref{Bow-Tie Cavity}). The
effect of the scanning magnet is minimized at the last point on the top
right corner of the resonator (see Figs. \ref{Bow-Tie Cavity}, \ref{High
Freq. Chaos Mode}). The electric field amplitude is zero on the boundaries,
and the pin perturbation does not shift the resonant frequency of the cavity
on the locations closest to the boundaries. Since this is the case, the
shift in the resonant frequency along the boundaries is purely a function of
the scanning magnet perturbation on the ferrite. Since the ferrite is placed
along the y-axis on left side of the cavity, the effect of the scanning
magnets is nearly constant as the magnet is moved in the y-direction. Using
this, we can subtract the difference between the last point in every y-axis
column and the last point on the top right corner from the data in that
column. This produces a flat background, leading to much cleaner images. The
image shown in Fig. \ref{High Freq. Chaos Mode} has had a background
subtracted, while those in the other figures have not.

\section{Limitations and Constraints}

We have several limitations on our measurements, including the resolution of
the images, resonant frequency to be imaged, and the perturbation size.
Eigenmode images are also affected by the perturbation caused by the
coupling probes. These issues are discussed in this section. Overall, the
limitations on the measurements are significantly reduced by the choice of
the perturbation and the method to scan the perturbation. Our imaging
technique works in a remarkably broad frequency range from 700 MHz to about
15 GHz, where the upper limit is imposed by the density of eigenenergies for
the cavity shown in Fig. \ref{Bow-Tie Cavity}.

\subsection{Spatial Resolution}

To improve the spatial resolution of the images, particularly at high
frequencies, we must use smaller perturbations. However, perturbations which
are too small yield frequency shifts comparable to the uncertainty in the
cavity resonant frequency due to noise (about 1 part in 10$^3$ of the 3 dB
bandwidth). Very small perturbations may also fail to follow the magnet
during the scan because they get stuck at the boundaries due to friction and
surface tension of the lubricant used in the cavity.

We observe that spheres with a diameter less than 3/32'' produce such a
small perturbation that the noise in the system is the dominant factor in
the eigenmode images. The smallest cylindrical pin which yields clear and
complete images has a diameter of 0.0085'' and a height of 0.1535''. This
pin is used for the 9 GHz to 15.4 GHz frequency range images shown in this
paper (see Fig. \ref{High Freq. Chaos Mode}).

\subsection{Mode Mixing}

Since the field strengths at a given point in the cavity are different for
different modes, shifts in the resonant frequencies due to a given
perturbation are different. This can result in resonant frequencies which
closely approach or cross over each other as the perturbation is scanned.
For higher frequencies, where the mode density is higher, smaller
perturbations must be used to avoid mode mixing.

Our technique for imaging the eigenfunctions of the hollow microwave cavity
shown in Fig. \ref{Bow-Tie Cavity} has high accuracy and success for
frequencies up to 15.4 GHz. This limit can be pushed up, for instance, by
using a silver layer on the inner walls of the cavity to increase the
quality factor of the resonant modes. This would make it possible to clearly
distinguish higher frequency modes from each other. In that case we can make
use of smaller perturbations since smaller shifts in the resonant frequency
of the cavity would be observable. For a typical bandwidth of 6.5 MHz for
resonant frequencies above 10 GHz, the mode separation has to be on the
order of 15 MHz to get reliable eigenmode images.

\subsection{Perturbation Movement}

Another limit on the spatial resolution of our images is imposed by the
perturbation movement mechanism. The stepper motors pull the magnet carriage
with belts, and this mechanism imposes a lower limit for step size. The
belts pulling the carriage carrying the magnets have backlash, and the
stepper motors occasionally miss steps. The smallest step size we can use is
0.05'' (using smaller step sizes will cause the scans to take more than 100
hours). Returning the x-y stage to its home position on the y-axis after
each scan line minimizes the errors due to backlash and the errors caused by
the missed steps.

The accuracy in perturbation location depends on the strength and the focus
of the magnetic field used to drag the perturbation, as well as the
separation of the magnet from the perturbation. The friction between the
perturbation and the cavity bottom surface while it is being dragged by the
magnet can have a significant effect on the image quality. The friction is
reduced by putting a thin layer of lubricant on the cavity floor. We find
experimentally that the approximate standard deviation in the location of
the pin is 0.025'', about half the smallest step size used in the
measurements (0.05'').

When the pin fails to follow the magnet, and either moves a little bit more
or less than the magnet, the pin tilts due to the high divergence of the DC
field of the scanner magnet. This tilt of the pin increases the magnetic
field contribution to the frequency shift. The tilt of the pin is due to
friction and can be minimized by moving it 0.1'' more than where it is
supposed to go and immediately drawing it back 0.1''. We have found that for
the pins with rough ends, the measured contribution to the frequency shift
of the magnetic field components can exceed 20\%. However for pins with
smooth ends, the measured magnetic field contribution is on the order of
10\%.

\subsection{Coupling Probe Perturbation}

The coupling ports perturb the cavity, resulting in a small change in the
eigenfunction patterns. It has been observed that the coupling probe
perturbation may even produce wave chaos in an integrable potential.\cite
{Haake} In order to measure the effect of the coupling on our images, a
single bow-tie resonator mode (Fig. \ref{Low Freq. Chaos Mode}) was imaged
at three different coupling strengths. The coupling strength was varied by
pulling the coupling probes out of the cavity. There was no visible change
in the structure of the eigenfunction images, although the signal-to-noise
ratio was degraded as the coupling was reduced.

\section{Conclusions}

Our technique has successfully imaged the standing wave patterns of
two-dimensional microwave resonators with and without time-reversal symmetry
based on the simple physical principles of cavity perturbation. This
technique allows us to image the eigenfunctions of a two-dimensional
microwave cavity analog of the Schr\"{o}dinger equation. Any hollow cavity
(where E$_z$ is not z-dependent) can be mounted to the scanning setup and
imaged. Since the magnet is scanned outside the cavity there is no need to
put holes on the cavity and change the boundaries other than at the coupling
ports. Our experimental setup allows us to make high quality, high
resolution images between 700 MHz and 15.4 GHz for a 1150 cm$^2$ area cavity.%
\cite{ColorPics}

Our images show that we have avoided mode mixing and reduced the RF magnetic
field contribution to an insignificant level. In addition, the data we have
obtained using this technique are in qualitative agreement with the
theoretical predictions made for the statistics of the field distribution.
The technique can be used in similar cases where the amplitude of the
electric field is the important quantity or eigenfunctions of the
Schr\"{o}dinger equation are needed for complicated quantum structures with
hard wall boundary conditions.

Acknowledgments:

We would like to acknowledge assistance from Paul So, Edward Ott, Allen
Smith and Tom Antonsen. This work was sponsored by the National Science
Foundation through an NSF NYI Grant DMR-92588183, and NSF DMR-9624021.

%\begin{figure}[tbp]
%\caption{Picture of a pin perturbation and the three components of the
%electromagnetic fields in transverse magnetic modes of thin microwave
%resonators.}
%\label{Pin Perturbation}
%\end{figure}
%
\begin{figure}[tbp]
\caption{Diagram of the quarter bow-tie resonator. The resonator is formed
by a 0.310'' deep pocket in the solid brass piece. For some experiments, a
ferrite bar is located on the left hand wall. Axes show the three components
of the electromagnetic fields in transverse magnetic modes of thin microwave
resonators.}
\label{Bow-Tie Cavity}
\end{figure}

\begin{figure}[tbp]
\caption{Schematic diagram of the experiment. Thick lines represent
microwave signal paths, while the thinner lines represent low frequency
control and data signal paths.}
\label{Schematic}
\end{figure}

\begin{figure}[tbp]
\caption{Close up schematic view of magnet ensemble and pin pertubation
inside the microwave cavity. Also shown are the coupling antennas in the
cavity.}
\label{Magnet Ensemble}
\end{figure}

\begin{figure}[tbp]
\caption{Distribution of frequency shifts for a) spherical perturbation and
b) pin perturbation in the resonant mode of a thin rectangular cavity shown
in Figs. 6 and 7.}
\label{Distributions}
\end{figure}

\begin{figure}[tbp]
\caption{a) Measured frequency shift versus pin perturbation location in a TM%
$_{170}$ mode of a rectangular microwave resonator. Also shown are b)
vertical and c) horizontal line cuts through the data and comparisons to the
theoretical frequency shifts.}
\label{Pin Image}
\end{figure}

\begin{figure}[tbp]
\caption{a) Measured frequency shift versus sphere perturbation location in
a TM$_{170}$ mode of a rectangular microwave resonator. Also shown are b)
vertical and c) horizontal line cuts through the data and comparisons to the
theoretical frequency shifts.}
\label{Ball Image}
\end{figure}

\begin{figure}[tbp]
\caption{Measured probability amplitude $|\Psi _n|^2 A$ in the quarter
bow-tie cavity, where A is the area of the entire cavity. This image is
obtained from a resonant mode at 3.46 GHz.}
\label{Low Freq. Chaos Mode}
\end{figure}

\begin{figure}[tbp]
\caption{Measured probability amplitude $|\Psi _n|^2 A$ in the quarter
bow-tie cavity, where A is the area of the entire cavity. This image is
obtained from a resonant mode at 5.37 GHz.}
\label{Mid Freq. Chaos Mode}
\end{figure}

\begin{figure}[tbp]
\caption{Measured probability amplitude $|\Psi _n|^2 A$ in the quarter
bow-tie cavity, where A is the area of the entire cavity. This image is
obtained from a resonant mode at 11.94 GHz. A ferrite bar is located along
the left wall, as shown in Fig. 2.}
\label{High Freq. Chaos Mode}
\end{figure}


\begin{references}
\bibitem{Slater}  L. C. Maier, Jr. and J. C. Slater, J. Appl. Phys. {\bf 23}%
, 68 (1952).

\bibitem{Rebeiz}  T. P. Budka, S. D. Waclawik, G. M. Rebeiz, IEEE Trans. MTT 
{\bf 44}, 2174 (1996).

\bibitem{Ashfaq}  A. Thanawalla, {\it et al}., submitted to Appl. Phys.
Lett. (1998).

\bibitem{Bloom}  A. S. Hou, F. Ho, D. M. Bloom, Electron. Lett. {\bf 28},
2302 (1992).

\bibitem{vdW}  D. W. van der Weide, and P. Neuzil, J. Vac. Sci. Technol. B 
{\bf 14}, 4144 (1996).

\bibitem{Anlage}  S. M. Anlage, {\it et al}., IEEE Trans. Appl. Supercon. 
{\bf 7}, 3686 (1997).

\bibitem{Sridhar1}  S. Sridhar, Phys. Rev. Lett. {\bf 67}, 785 (1991).

\bibitem{Sridhar2}  S. Sridhar, D. O. Hogenboom, B. A. Willemsen, J. Stat.
Phys. {\bf 68}, 239 (1992).

\bibitem{Stein+Stockmann}  J. Stein and H. -J. St\"{o}ckmann, Phys. Rev.
Lett. {\bf 68}, 2867 (1992).

\bibitem{Wu}  D. H. Wu, Ali Gokirmak, J. S. A. Bridgewater, and S. M.
Anlage, submitted to Phys. Rev. Lett.

\bibitem{Hupert1}  Julius J. Hupert and Gary Ott, Am. J. Phys. {\bf 34}, 260
(1966).

\bibitem{Hupert2}  Julius J. Hupert , IRE Trans. Circuit Theory {\bf XX},
425 (1962).

\bibitem{PaulSo}  Paul So, Steven M. Anlage, Edward Ott, and Robert N.
Oerter, Phys. Rev. Lett. {\bf 74}, 2662 (1995).

\bibitem{Antonsen}  T. M. Antonsen, Jr., E. Ott, Q. Chen, and R. N. Oerter,
Phys. Rev. E {\bf 51}, 111 (1995).

\bibitem{Heller}  Eric J. Heller, Patrick W. O'Connor, and John Gehlen,
Physica Scripta {\bf 40}, 354 (1989).

\bibitem{Arrick}  The scanning system consists of a XY-30 scanner, and the
MD-2b dual stepper motor driver obtained from Arrick Robotics of Hurst,
Texas.

\bibitem{TransTech}  The ferrite was obtained from TransTech of Adamstown,
Maryland. It is model \# CVG1850, with a saturation magnetization of 1850
Oe, a resonance linewidth of 15 Oe, a dielectric constant of approximately
14.8, and dielectric loss tangent less than 0.0002.

\bibitem{PaulSoPhD}  Paul So, Ph.D. Thesis, University of Maryland, 1996.

\bibitem{ColorPics}  Color images of these and other wave chaotic
eigenfunctions are available at
http://www.csr.umd.edu/research/hifreq/mw\_cav.html.

\bibitem{Prigodin}  V. N. Prigodin, {\it et al}., Phys. Rev. Lett. {\bf 75},
2392 (1995).

\bibitem{Haake}  F. Haake, G. Lenz, P. Seba, J. Stein, H.-J. Stockmann and
K. Zyczkowski, Phys. Rev. A {\bf 44}, R6161 (1991).
\end{references}
\end{document}